\documentclass[english]{achemso}
\usepackage[T1]{fontenc}
\usepackage{amstext}
\usepackage{graphicx}
\usepackage{subscript}
\usepackage{gensymb}

\makeatletter

\title{Ferroelectric Fractals: Switching Mechanism of Wurtzite AlN}

\author{Drew Behrendt}
\affiliation{Department of Chemistry, University of Pennsylvania, Philadelphia, Pennsylvania, 19104-6323, USA}
\author{Atanu Samanta}
\affiliation{Department of Chemistry, University of Pennsylvania, Philadelphia, Pennsylvania, 19104-6323, USA}
\author{Andrew M. Rappe}
\affiliation{Department of Chemistry, University of Pennsylvania, Philadelphia, Pennsylvania, 19104-6323, USA}
\email{rappe@sas.upenn.edu}

\makeatother

\usepackage{babel}
\begin{document}

\section*{Abstract:}
The advent of wurtzite ferroelectrics is enabling a new generation of ferroelectric devices for computer memory that has the potential to bypass the von Neumann bottleneck, due to their robust polarization and silicon compatibility. However, the microscopic switching mechanism of wurtzites is still undetermined due to the limitations of density functional theory simulation size and experimental temporal and spatial resolution. Thus, physics-informed materials engineering to reduce coercive field and breakdown in these devices has been limited. Here, the atomistic mechanism of domain wall migration and domain growth in wurtzites is uncovered using molecular dynamics and Monte Carlo simulations of aluminum nitride. We reveal the anomalous switching mechanism of fast 1D single columns of atoms propagating from a slow-moving 2D fractal-like domain wall. We find that the critical nucleus in wurtzites is a single aluminum ion that breaks its bond with one nitrogen and bonds to another nitrogen; this creates a cascade that only flips atoms directly in the same column, due to the extreme locality (sharpness) of the domain walls in wurtzites. We further show how the fractal shape of the domain wall in the 2D plane breaks assumptions in the KAI model and leads to the anomalously fast switching in wurtzite structured ferroelectrics.

\section*{Introduction:}

Ferroelectric materials are non-centrosymmetric insulating crystals with a polar axis that can be switched by an external electric field. Ferroelectrics have been used in functional devices such as random access memory and tunable capacitors since the early 1950s.\cite{Kim23-FEmem,Song2021} These devices leverage the unique piezoelectric, pyroelectric, and ferroelectric properties of the chosen material, so the device architecture is highly dependent on material choice. For most of these devices, perovskites such as BaTiO$_3$, PbTiO$_3$, Pb(Zr,Ti)O$_3$, or Pb(Mg$_{1/3}$Nb$_{2/3}$)O$_3$-PbTiO$_3$ are the prototypical ferroelectric of choice.\cite{GAO20201,Kim2023-wzolsson,Qi2015} These materials have high polarization, low coercive field, high dielectric constant, and they have been manufactured for decades. However, perovskite oxides must be rather thick to retain polarization, and perovskites cannot be grown well at sufficiently low temperature to be compatible with silicon, which is necessary for back end of the line compatibility with computer chips. \cite{GAO20201}

\begin{figure}
    \centering
    \includegraphics[width=1\linewidth]{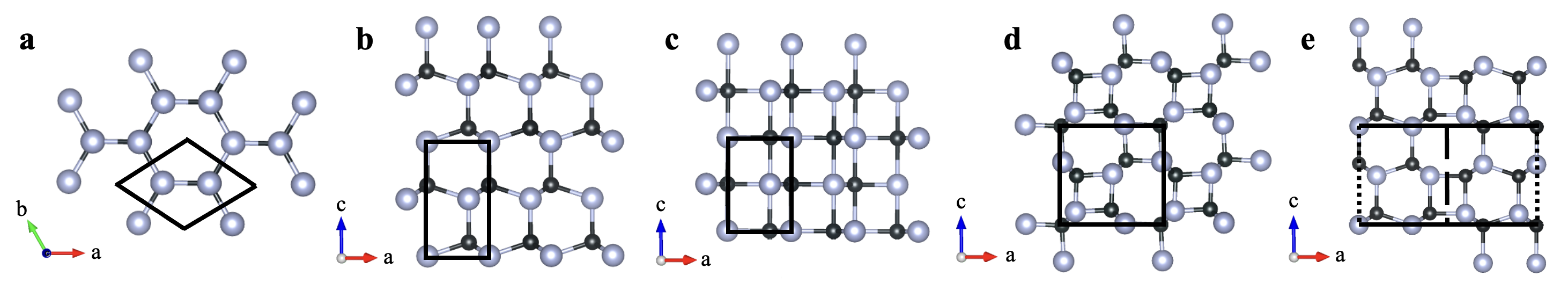}
    \caption{Important metastable bulk structures for AlN switching. Black atoms are Al and grey are N. All structures have the same hexagonal lattice viewed along the c-axis (a) and can all be accessed from each other through atomic displacements along the c-axis. (b) is the stable bulk wurtzite phase. (c) is the hexagonal boron-nitride phase that is the non-polar reference structure. (d) is an anti-ferroelectric (AFE) phase observed in $\beta$-BeO. (e) Shows a planar 180\degree \vspace{1mm} domain wall that is between wurtzite phases of opposite polarity away from the wall and a $\beta$-BeO type structure at the interface. Black rectangles show the repeating bulk unit, with a thick dashed line highlighting the domain wall and thin dashes representing repeating of either the domain wall supercell or bulk-like wurtzite.}
    \label{fig:struct}
\end{figure}

Recently, ferroelectricity has been discovered in many polar materials where polarization reversal has been unlocked by doping historically unswitchable polar materials, such as wurtzite boron- or scandium-doped aluminum nitride (Fig. \ref{fig:struct}b). \cite{Ferri21-FEeverywhere,Fichtner2019,Hayden2021,Liu2023,Wang22} This advance provides new classes of materials offering favorable properties for next-generation devices. For example, wurtzite ferroelectrics retain polarization at high temperatures with no measurable Curie temperature.\cite{Drury22} Wurtzites also provide high thermal conductivity and great silicon compatibility; thus, devices based on wurtzites can remain functional in high temperature environments and can be easily integrated with electronic devices.\cite{Song2021,Kim2023-wzolsson} The main drawback of this class of ferroelectric is the extremely high coercive field (>3 MV/cm for even the thinnest films) which is extremely close to the breakdown field. \cite{yazawa21} High coercive field leads to a higher energy cost to switch and more facile breakdown than other ferroelectrics. However, these problems can potentially be improved with materials engineering and optimization once the microscopic switching mechanism is fully understood. 

Historically, the Johnson Mehl Avrami Kolmogorov (JMAK) equations have been used to model the kinetics of crystallization via nucleation and growth.\cite{Avrami39,Blazquez22} Later, the Kolmogorov Avrami Ishibashi (KAI) model was introduced to expand this theory to ferroelectric switching.\cite{Ishibashi71} In this model, the flipped fraction $f$ is represented by $f(E,T,t)=1-e^{X(E,T,t)}$ where $X(E,T,t)$ is the extended transformed fraction that depends on the strength of the electric field, temperature, and time. The value of $X(E,T,t)=-(t/t_0)^n$, where $n$ is one more than the dimension of growth when including nucleation.\cite{Ishibashi71,Blazquez22} For a 3D material with a constant domain nucleation rate, for example, $n$ should equal 4. It has been reported that for wurtzite AlN doped with scandium or boron, the experimental switching kinetics can be fitted to the KAI model, with $n$ equal to 7 and 11, respectively, in stark contrast to other known ferroelectrics. \cite{Yazawa2023-kai,Lu24}

Due to the novelty of ferroelectricity in this class of materials, there has yet to be offered a complete microscopic atomic mechanism for how the switching occurs and why the KAI model is violated. This current gap in fundamental understanding hampers efforts to engineer these materials atomistically to improve the desired response and ultimately lower coercive field. Previous studies based on density functional theory (DFT) and TEM experimentation have proposed multiple possible pathways for homogeneous switching through a bulk intermediate phase such as h-BN or $\beta$-BeO (Fig. \ref{fig:struct}c and d).\cite{Calderon2023,Huang2022,Krishnamoorthy_2021,Wolff24,Yassine24} While these structures certainly represent non-polar intermediate states that can be visited locally during macroscopic switching, the DFT calculations are inherently limited to investigating simulation sizes on the order of 100 atoms. While they provide interesting insights, these simulations are incapable of investigating mechanisms for heterogeneous ferroelectric domain nucleation and growth. 

To move beyond DFT to computationally model emergent macroscopic phenomena, classical molecular dynamics (MD) is an enabling tool. While there is a rich history of domain nucleation and growth simulations in ferroelectric perovskites with MD, wurtzites have not been extensively studied with MD. \cite{Shin2007,Liu2016,Qi2015,Samanta2022} To generate an interatomic potential that can handle $10^5-10^6$ atom supercells yet still approach the accuracy of DFT, one could fit either an empirical potential, a model Hamiltonian, or an artificial neural network (ANN), to DFT snapshots of the material of interest. Here, we use our recently developed machine learned force field (MLFF) for bulk aluminum nitride to uncover the macroscopic switching mechanism for wurtzites.\cite{Behrendt24} We use these insights to explain some recently observed experimental anomalies that show the uniqueness of this new ferroelectric material. We hope that the deepened understanding of this mechanism will lead to new materials engineering to power the next generation of integrated and high-temperature ferroelectric devices.

In our previous work, we confirmed accurate behavior for our MLFF relative to DFT and experiment for the energetics and phonon frequencies of the major bulk phases of AlN and with domain walls under $NPT$ simulation conditions.\cite{Behrendt24} Crucially, our work and other recent theoretical studies have shown that the stable uncharged 180$^\circ$ domain walls in wurtzite ferroelectrics has the structure of the metastable $\beta$-BeO (Fig. \ref{fig:struct}d) phase at the boundary, allowing complete polarization reversal to occur between two adjacent unit cells (Fig. \ref{fig:struct}e).\cite{Calderon2023,Liu2023,Herng2012,Huang2022} We also found very little interaction between uncharged domain walls and a nearly constant wall energy as a function of their density. Here, we investigate the dynamical properties of domain wall migration.

\section*{Results and Discussion:}

\begin{figure}
    \centering
    \includegraphics[width=1\linewidth]{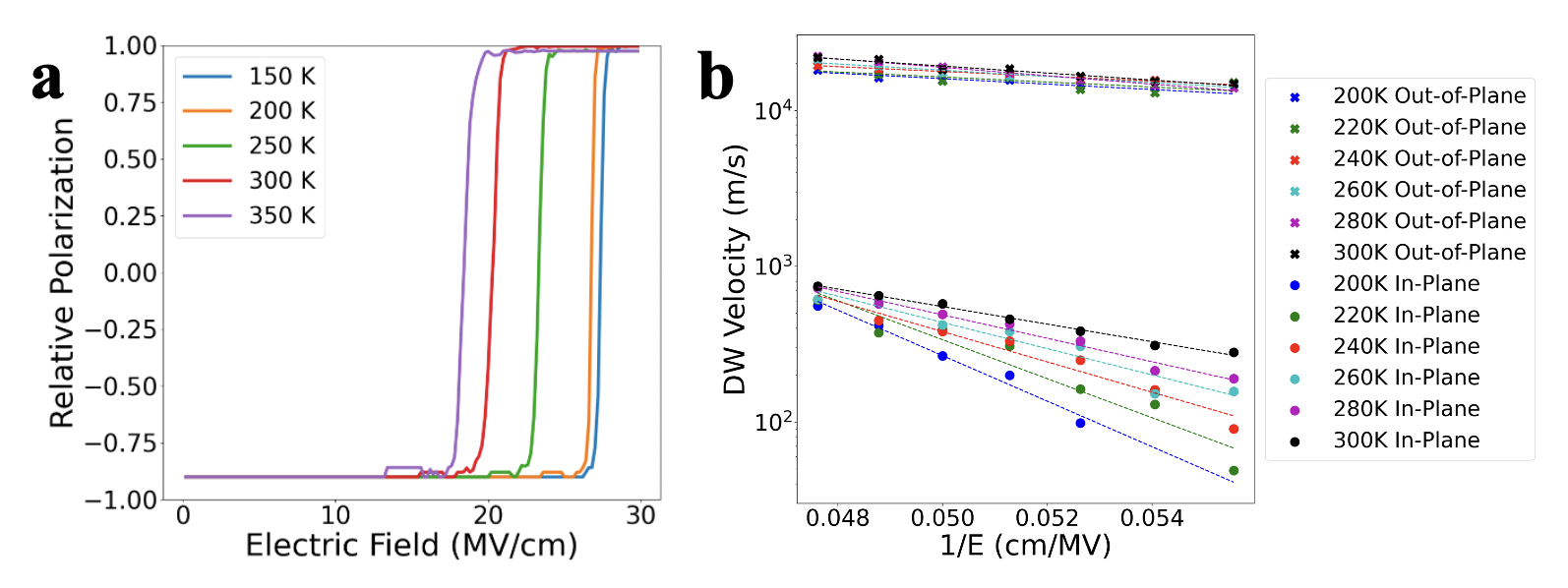}
    \caption{a, Half hysteresis loops for pre-nucleated bulk AlN from molecular dynamics simulations. The field was slowly increased from zero to 30 MV/cm over 30 ns with a step size of 1 fs. b, Merz law fittings of domain wall velocities along both polar and nonpolar axes as a function of temperature.}
    \label{fig:quant}
\end{figure}

To compare our simulations to real materials, in Fig. \ref{fig:quant}a, we show the characteristic trends of extremely square hysteresis loops due to very fast switching events as well as a direct relationship between increased temperature and lowered coercive field are captured by our simulations, in agreement with recent experiments.\cite{Zhu2021,Yazawa2023-kai,Yazawa2023-wakeup} Similar to experiment, these results were obtained by increasing the $E$ field strength and calculating the polarization at each time step, showing that our simulations correctly reflect the macroscopic switching behavior of wurtzite ferroelectrics, which is uniquely enabled by our large scale simulations. \cite{Hayden2021,Yazawa2023-kai} Additionally, using the formula $E_c(T)=E_c^0e^{-\nu/k_BT}$ and our results from Fig. \ref{fig:quant}a, we fit a pseudo activation energy, $\nu$, from the temperature dependence of coercive field to be 9.4 meV, in relative agreement with experimental values for ZnMgO and AlBN of 20-40 meV.\cite{Casamento24} This indicates a low energy switching barrier from a pathway that must be heterogeneous in nature and governed by domain nucleation and growth rather than homogeneous bulk switching. 

Next, we quantify the switching speed of the proposed domain wall motion that governs the switching, including the anisotropy between in-plane domain wall motion along the non-polar axes and out-of-plane wall motion along the polar axis. We calculated average switching velocities along the 2D (in-plane) in plane direction and along the 1D lines (out-of-plane). The velocities for a range of temperatures and fields are plotted in Fig. \ref{fig:quant}b. We find that the results are well fit with Merz's law ($v \propto e^{-E_a/E}$) for both charged and uncharged domain wall motion, with a massive velocity anisotropy. Additionally, fitting the slopes of Fig. \ref{fig:quant}b to Merz's law, we find an activation field for in plane switching to be 130 and 240 MV/cm at 300 K and 200 K, respectively, while for out of plane switching the value is only $\approx$50 MV/cm for both temperatures.

\begin{figure}
    \centering
    \includegraphics[width=1\linewidth]{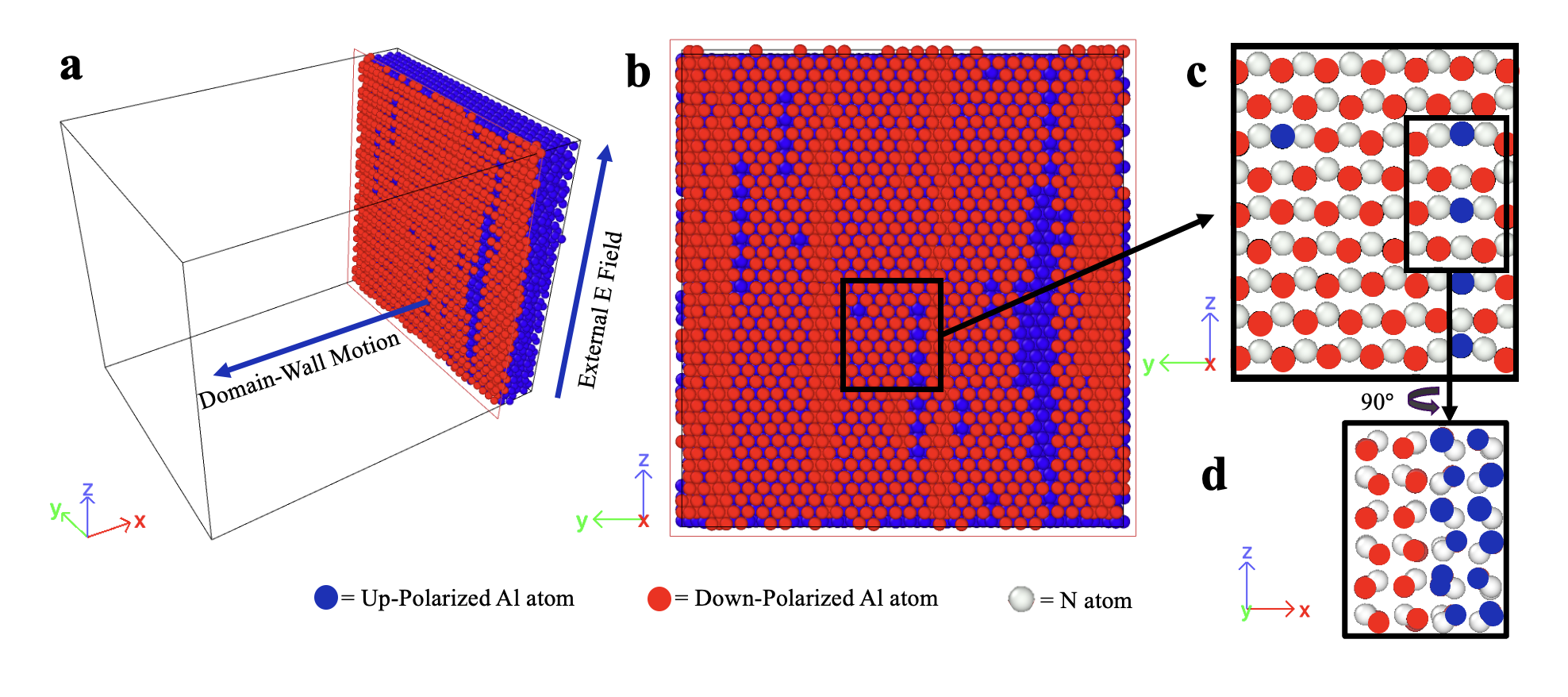}
    \caption{a, View of entire simulation cell nucleated from a flat domain wall. b, Nucleation event in AlN along a [100] cut of the bulk sample. In panels a and b, only Al atoms are shown for clarity. Blue and red Al atoms represents polarization down and up atoms, respectively, with the field pointed up. An atom turns from red to blue when it becomes closer to the nitrogen above than the one below. Single blue atoms are the critical nuclei at this level of field, with a 50-50 chance of either nucleating a new column of atoms flipping to blue or returning to red. c, Inset of a microscopic snapshot at the nucleation point with both Al (blue and red) and N (grey) atoms shown. In c, only two planes of atoms are shown to see the local bonding as a new line of atoms nucleates and the domain wall grows. d, Rotated subsection of inset to show $\beta$-BeO-like domain wall boundary.}
    \label{fig:nuc}
\end{figure}

Because we observe the experimental switching behavior in a computational model with no dopants, vacancies, or surface effects, we posit that the key aspects of the ferroelectric switching mechanism in this work are due to the material symmetry and bonding characteristics shared by all wurtzites. Thus, to find the proposed universal size and shape of the nucleation events, we view a 2D cut of the material as the domain propagates in Figure \ref{fig:nuc}, allowing us to see the microscopic origins of domain growth. 

Through these simulations, we found that each individual growth event adjacent to the domain wall starts by a single Al ion breaking its axial N-bond, in turn nucleating an individual 1D column of flipped atoms. This is vastly different from typical perovskite ferroelectrics, where the growth happens through a diffuse square critical nucleus of many unit cells that must flip simultaneously in order to survive and grow.\cite{Shin2007} Each line is formed by alternating Al and N atoms that must both displace equally in order to switch polarization. The domain walls are sharp, completely reversing between adjacent lines. We report that this sharp reversal occurs between any of the three neighboring lines adjacent to a nucleated line and not just in a uniformly flat domain wall (Fig. \ref{fig:struct}a, e). Fig. \ref{fig:nuc}d shows that the local $\beta$-BeO interface is formed when a new line nucleates, but it occurs only between the unflipped neighbors and the newly flipped line.

Additionally, we see that the tetrahedron of bonds surrounding each Al and N atom is distorted when, and only when, it lies along the domain boundary. As seen along the domain boundary in Fig. \ref{fig:struct}e and Fig. \ref{fig:nuc}c, the atoms that are on the boundary have a 90$^\circ$ angle between the c-axis bond and the bond to the flipped line and a much larger bond angle between the atoms along the a-axis. This distortion leads to a lower barrier for flipping when, and only when, a column is next to a domain wall and is the reason that only the directly adjacent lines are affected by a flipped neighbor. 

From the Merz law fittings of activation fields and visualizing the switching mechanism, we see that there are orders of magnitude differences in the growth rates along the polar and nonpolar axes. The charged domain wall that forms during nucleation of a new line is very high in energy, and the rest of the 1D line will quickly switch to align with the field and the switched part of the line with a low energy barrier. However, the uncharged domain walls are more stable and only affect new lines that border the domain wall. Thus, the switching within the 2D non-polar plane can be effectively separated from the 1D switching along the polar axis. Each nucleated line switches so fast that the domain growth can be viewed as a 2D plane of flipped lines, and the growth rate is limited by the flipping rate in this plane. 

\begin{figure}
    \centering
    \includegraphics[width=1\linewidth]{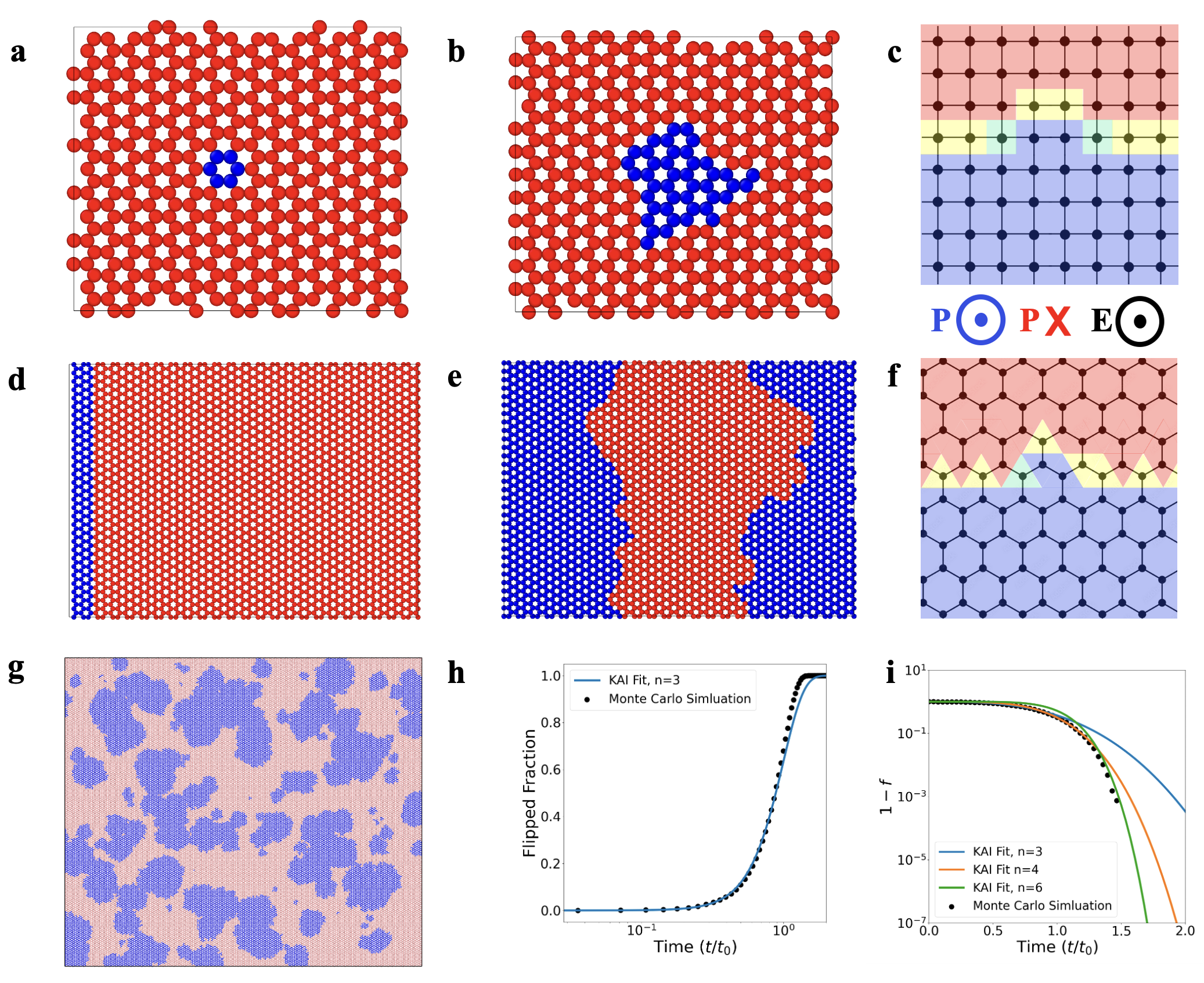}
    \caption{Two-dimensional domain growth in bulk AlN captured by a slice along the [001] direction, with blue and red atoms representing polarization out of and into the page, respectively. a, the minimum size stable domain in AlN which evolves to b  after 2 ps of applied field. d, A flat domain wall that evolves to e after 15 ps of applied field. d,e show the simulation cell size used for domain wall velocity calculations. c, Pictorial example for a flat domain wall on a square lattice; green and yellow spots represent cells having 2 and 1 flipped neighbors, respectively. f, Pictorial example for a flat domain wall on a hexagonal lattice; no two units on the flat domain wall are directly in contact with each other, and green spots only can appear after two independent flipping events which are not constrained to happen along the flat domain wall. Flipping of one column of atoms on a flat domain wall on a hexagonal lattice will not affect other atoms along the wall until randomly near each other, leading to the triangular snowflake pattern observed in simulations. g, Image from Monte Carlo simulations used to model mesoscale 2D growth. h, Comparison between Monte Carlo simulation results and a KAI model fitted to the respective geometry. i, Extrapolated KAI fitting of $X(t)$ vs t to show the mesoscale dependence of $f$ on $-t^n$, particularly as the switching event finishes.}
    \label{fig:2Dmov}
\end{figure}

To see clearly the in-plane domain wall migration patterns, snapshots from a 2D plane of AlN during a MD switching simulation are shown in Figure \ref{fig:2Dmov}. From these simulations, we see that three major structural properties of wurtzites lead to the domain wall roughness that emerges in the 2D grid of Fig. \ref{fig:2Dmov}: the high stability of the uncharged domain walls, the single unit cell interaction range of each switched line only affecting its nearest neighbors, and the hexagonal symmetry, leading to each line only having three neighbors. Because of this, when the system is initialized with a flat domain wall, no two unflipped neighbors along the wall directly interact with each other and will flip independently. This independent line growth that is only affected by the polarization of the three nearest neighbors ultimately leads to random snowflake-like 2D patterns that emerge when ferroelectric domains in a large supercell are allowed to grow. The rate of flipping for an individual line is dependent on how many of its neighboring three lines are also flipped. When no neighbors are flipped, it is very energetically unfavorable to break a tetrahedron and initiate flipping independently. When one, two, or three neighboring lines are flipped, however, the distorted local environment ($\beta$-BeO-like structure) allows for easier breaking and reforming of axial nitrogen-aluminum bonds to realign with the external field. More flipped neighbors cause a more distorted local tetrahedron and thus easier (and quicker) flipping. 

We also would like to note that the proposed mechanism is for a simulation of bulk AlN with a uniform electric field, and may be realized differently for a real material with electrode and device effects. For example, given that there will be dielectric screening of the electric field in a real device, it is reasonable to expect that domains nucleation and growth would initiate near the electrode instead of randomly throughout the material. This would not necessarily alter the kinetics of switching, but could result in the formation of spiked domains that propagate from the electrode throughout the rest of the material. Additionally, as the growth is limited in the 2D plane, we expect similar switching kinetics regardless of where individual columns of atoms are nucleated. Furthermore, while we find that head-to-head charged domain walls are incredibly unfavorable in our pure bulk simulations, it would also be reasonable to assume that this could be stabilized in experimentally observed leaky thin films that can passivate this extra charge, leading to more gradual domains.\cite{Wolff24}

Furthermore, our MD simulations only allow for studies of domain growth from a previously formed domain, because initial formation of nuclei is very high energy and rare in pure AlN with no dopants. Therefore, in order to capture the dynamics as a function of domain nucleation rate, we performed Monte Carlo simulations on a 2D hexagonal grid. Each point on the grid is assigned a probability to flip based on the number of flipped neighbors, with a small probability to form an initial domain and a much higher probability to flip when more neighboring points are already flipped. This leads to a controllable linear nucleation rate and a linear growth rate in the $x$ and $y$ directions. The results for the simulation of a moderate growth rate is shown are Fig. \ref{fig:2Dmov} g,h,i.

The primary consequence of this localized flipping of single columns of atoms is that the domain walls no longer remain convex as they propagate. Measuring the perimeter of a smooth curve or wall is a simple practice of using successively smaller line segments to approximate the perimeter, the limit of which is the true line length as the segment approaches zero. This test fails, however, for fractal structures like the famous examples of the Koch Snowflake or the perimeter of a coastline. The degree to which these structures deviate from the perfect line or curve can be captured in the fractal dimension, which relates the change in perimeter to the length of segment used to measure it or by comparing the relative perimeter and area of the fractal regions.\cite{Husain2021,Florio19,Catalan08,Mandelbrot67} Using this method and the intrinsic perimeter and area of the observed domains, we can approximate the fractal dimension ($d\approx log(A)/log(P)$) of the growth using traces of the domains in Fig \ref{fig:2Dmov}g and other snapshots of growth from the MC simulations to be $\approx1.34$. Similarly, if we apply our method to experimental images published by Guido et al., we find a fractal dimension of $\approx1.29$.\cite{Guido24} While the fractal nature of macro-scale ferroelectric domains has been documented by many other works for perovskites and other ferroelectrics,\cite{Moroz20,Scott07,Catalan08,Everhardt19} here we show that this fractal dimension extends all the way down to the atomic scale, and is not purely caused by uneven coalescence of domains.

The time evolution of polarization switching for traditional ferroelectrics, even those that are macroscopically fractal, is well modelled by the KAI (or KJMA) model. However, this model is not fully applicable for the microscopically fractal nitrides at high fields, with $n$ values reported up to 11.\cite{Yazawa2023-kai} Previously, this anomalously fast switching was interpreted as indicating a time-dependent nucleation rate, with nuclei forming faster as flipping proceeds, in order to fit the model to experimental data.\cite{Yazawa2023-kai,Guido24} However, in Fig. \ref{fig:2Dmov} h,i we show that the same deviation seen in experiments from conventional KAI ($n$ should be 3 for this 2D grid) is seen in our simulations and thus can be accounted for as an intrinsic feature of our proposed growth mechanism with constant nucleation and fractal domain wall evolution.  

This wurtzite domain reversal mechanism leads to a breaking of the conventional KAI model because that it violates a fundamental assumption of the model, that the domains considered in the growth are convex and approximately oval in shape. For prototypical perovskite ferroelectrics, this model can allow for large scale domains to appear fractal, as uniform combinations of circular domains can appear uneven on a large enough scale. However, because of the extreme locality of the domain walls and individual lines propagating from the domain walls in wurtzites like AlN, the growth in the 2D nonpolar plane is microscopically fractal in nature and not smooth or convex. Therefore, when the switching mechanism is dominated by growth, the domain walls become more fractal and have a higher perimeter to area ratio than assumed by the KAI model. Thus, domains initially grow similarly as predicted from conventional KAI. As domains coalesce, fractal domain walls are eliminated, showing faster growth than expected at the later stages. The increased perimeter of the fractal domains causes larger areas of domain coalescence than typical uniform growth and thus greater deviation from the KAI model due to the large areas exposed to the increased flipping rate at the end of the switching process.

When we raise the nucleation rate in our simulations sufficiently, we find agreement with the conventional KAI model, because the formed domains do not have enough time and space to develop significant fractal character without overlapping, and the assumptions of the KAI model approximately hold. However, when the nucleation rate is lower relative to the growth rate, the switching deviates significantly from the conventional KAI prediction because the process is dominated by the fractal growth, without domain coalescence until later in the switching process. 

\section*{Conclusion:}

Here we present a new multi-scale switching mechanism for pure wurtzite phase bulk aluminum nitride. This analysis was gleaned through observation of molecular dynamics simulations built upon a machine learned force field for AlN. Instead of the homogeneous switching pathways that have been recently studied with DFT, we propose a new pathway where individual lines of atoms nucleate along the domain wall and switch quickly along the polar direction with the field while forming complex fractal snowflake-like hexagonal patterns in the nonpolar plane. This hundred-fold anisotropy of domain propagation, the extreme locality of the domain wall width, and the hexagonal connectivity lead to our proposed mechanism of 1D lines of atoms flipping to form a 2D fractal domain wall. The resulting Monte Carlo simulations built on this model show that the anomalous switching kinetics reported experimentally is intrinsic to the breaking of the conventional KAI model assumptions by the features of this fractal mechanism and do not require invocation of time-dependent nucleation rates. We hope that the present analysis serves to provide further explanation of experimental anomalies found in wurtzite switching and will ultimately help to advance materials engineering based on leveraging this pathway to enable new memory devices built with nm scale on-chip ferroelectrics.

\section*{Methods:}

In this work we consider pure bulk AlN with no dopants or vacancies to model fundamental properties of switching in wurtzites. The machine learned force field (MLFF) potential was developed using the AENET program\cite{Artrith2016} and was rigorously tested in our previous work. It is important to note that our potential was trained not only with a wide variety of meta-stable phases and structures, but also with images generated with NEB structural states along the homogeneous switching pathways. Here, we use the AENET-LAMMPS\cite{Chen2021} addition to the typical LAMMPS program \cite{LAMMPS}) for MD simulations. Electric field is applied by uniformly applying force to atoms along the c-axis proportional to their Born effective charges. The Born Effective charge tensor is calculated for bulk AlN with Quantum Espresso \cite{Giannozzi_2009} and norm-conserving optimized pseudopotentials generated with OPIUM.\cite{Rappe90,Rappe99} 

Experimentally, AlN has been shown to break down at high $E$ field due to the proximity of the coercive voltage to the breakdown voltage.\cite{Hayden2021} In the present simulations, there is no breakdown from excited electrons in the system or other forces, so the applied field can be increased to accommodate the short time frames of MD simulations. 

Switching simulations with the MLFF were conducted with a $24\times30\times40$ supercell of AlN with 115200 atoms and 1 fs timesteps. All simulations were equilibrated for 10 ps before $E$ field was applied and data was collected. The considered cell is shown in Fig. \ref{fig:nuc}a. Domain wall velocity simulations were collected over 15 ps, with the atomic positions of each atom collected every 200 fs. The polarization of each Al atom was calculated by its displacement from the center of mass between the nearest two N atoms in the polar axis. As the N and Al atoms displace equally during the switching, only displacement of Al atoms was considered for consistency. 

The Monte Carlo code was written in-house using Python to replicate the 2D switching patterns observed via molecular dynamics. Probabilities for each atom to flip from state 0 to 1 at a given time were assigned based on the state of the three nearest neighbors at the previous step, with probabilities increasing from 0 to 3 flipped neighbors. A higher field in these simulations corresponds to higher flipping probabilities. All visualizations of domains were generated with the OVITO program. \cite{ovito}

\clearpage 

\section*{Corresponding Author:}

{*}E-mail: rappe@sas.upenn.edu

\section*{ORCID:}

Drew Behrendt: 0000-0003-4701-2722; Atanu Samanta: 0000-0001-7918-2772; Andrew M. Rappe: 0000-0003-4620-6496

\section*{Acknowledgements:}
This material is supported by the U. S. Department of Energy, Office of Science, Office of Basic Energy Sciences Energy Frontier Research Centers program under Award Number DE-SC0021118. Computational support was provided by the High-Performance Computing Modernization Office of the Department of Defense and the National Energy Research Scientific Computing Center (NERSC), a U.S. Department of Energy, Office of Science User Facility located at Lawrence Berkeley National Laboratory, operated under Contract No. DE-AC02-05CH11231. 

\clearpage

\bibliography{AlNmech}

\end{document}